\documentstyle[preprint,aps]{revtex}
\begin{document}
\draft
\title{Cosmological Perturbations of Quantum-Mechanical Origin
and Anisotropy of the Microwave Background}
\author{L. P. Grishchuk}
\address{McDonnell Center for the Space Sciences, Physics Department}
\address{Washington University, St.Louis MO 63130}
\address{and}
\address{Sternberg Astronomical Institute, Moscow University}
\address{119899 Moscow, V-234, Russia}
\maketitle
\begin{abstract}
Cosmological perturbations generated quantum-mechanically (as a particular
case, during inflation) possess statistical properties of squeezed quantum
states. The power spectra of the perturbations are modulated and the angular
distribution of the produced temperature fluctuations of the CMBR is quite
specific.  An exact formula
is derived for the angular correlation function of the temperature
fluctuations caused by squeezed gravitational waves.  The predicted
angular pattern can, in principle, be revealed by the COBE-type observations.
\end{abstract}
\newpage

The recent discovery by COBE~[1] of the angular variations of CMBR
makes it necessary to analyze in greater detail the observational
consequences of the quantum-mechanical generation of cosmological
perturbations. The underlying physical reason for the generating process
is the parametric (superadiabatic) amplification of classical perturbations
and the associated quantum-mechanical particle pair creation in the variable
gravitational field of the homogeneous isotropic Universe. As a result of
the parametric coupling between the quantized perturbations and the variable
classical ``pump'' field, the initial vacuum state of the perturbations
evolves (in the ${\rm Schr\ddot o dinger}$ picture) into a strongly
squeezed vacuum
state possessing very specific statistical properties. The generated
fluctuations can be viewed, classically, as a stochastic collection of
standing waves. The mechanism itself and its main results concerning
squeezing are valid for gravitational waves and progenitors of density
perturbations~[2,3]. A particular variable gravitational field,
that may be responsible for the amplification process, is provided by one or
another type of the inflationary expansion. It is often stated that inflation
generates ``Gaussian perturbations with randomly distributed phases''.
However, this is not the case: the phases of all modes of
perturbations are essentially constant and fixed~[3] which leads to
standing waves, modulated spectra of the generated perturbations and a specific
angular distribution of the temperature fluctuations of CMBR over the sky, as
will be shown below.

In this paper we will analyze, mostly, gravitational waves.  For our
purposes it is sufficient to consider perturbations in a spatially-flat
FLRW universe \\
$ds^2=a^2( \eta )(d \eta^2 -dx^2 - dy^2 -dz^2 )$
where $a(\eta)$ is the cosmological scale factor.

The quantum-mechanical operator for the gravitational-wave field
can be written in the general form
\begin{equation}
h_{ij}(\eta , {\bf x})=C \int_{-\infty}^\infty d^3 {\bf n}
\sum_{s=1}^2 p_{ij}^s ({\bf n}) [a_{\bf n}^s (\eta ) e^{i{\bf nx}}
+a_{\bf n}^{s+} (\eta )e^{-i{\bf nx}} ]
\end{equation}
where $C$ is a constant combining all the numerical coefficients,
$p_{ij}^s({\bf n})$ are two $(s=1,2)$ polarization tensors and
$a_{\bf n}^s(\eta)$, $a_{\bf n}^{s+}(\eta)$,
are (Heisenberg) operators for each mode $\bf n$ and for each
polarization state $s$.

The polarization tensors $p_{ij}^s(\bf n )$ satisfy the
``transverse-traceless'' conditions $p_{ij}^s n^j=0$, $p_{ij}^s\delta^{ij}=0$
and leave independent only two components of $h_{ij}$ for each ${\bf n}$-mode
of the field.  For a wave travelling in the direction \\
${\bf n}/n=(\sin \theta\cos\varphi ,\sin\theta\sin\varphi ,\cos\theta )$
the polarization tensors are \\
$p_{ij}^1({\bf n})= l_i l_j -m_i m_j$,
$p_{ij}^2({\bf n})= l_i m_j +l_j m_i$, where $l_j$, $m_j$ are two unit
vectors orthogonal to ${\bf n}$
and to each other:  $l_j=(\sin\varphi ,-\cos\varphi ,0)$,
$m_j=(\cos\theta\cos\varphi\, ,\cos\theta\sin\varphi\, ,-\sin\theta )$
for
$\theta < \pi /2$ and
$m_j=-(\cos\theta\cos\varphi ,\cos\theta\sin\varphi ,-\sin \theta$)
for $\theta > \pi /2$.

The operators
$a_{\bf n}^s(\eta )$, $a_{\bf n}^{s+}(\eta )$
are annihilation and creation operators for waves (particles)
travelling in the direction ${\bf n}$.  The time evolution of
$a_{\bf n}^s(\eta )$, $a_{\bf n}^{s+}(\eta )$
is governed by the Heisenberg equations of motion for each mode
${\bf n}$ and for each polarization state $s$
(index $s$ is omitted here but will be restored later):
$da_{\bf n}/d\eta=-i[a_{\bf n},H],$
$da_{\bf n}^+/d\eta=-i[a_{\bf n}^+,H]$.
The Hamiltonian $H$ to be used in these equations has the form
$H=na_{\bf n}^+ a_{\bf n} + na_{-{\bf n}}^+ a_{-{\bf n}}
+ 2\sigma (\eta ) a_{\bf n}^+ a_{-{\bf n}}^+
+ 2\sigma^{\ast} (\eta ) a_{{\bf n}} a_{-{\bf n}}$
where the coupling function $\sigma (\eta )=ia^\prime /2a$
and ${}^\prime =d/d\eta$.  The solution to the Heisenberg
equations of motion can be written as
\begin{equation}
    a_{\bf n}(\eta ) = u_n(\eta )a_{\bf n} (0)
  + v_n(\eta )a_{-{\bf n}}^+ (0) \, , \quad
    a_{\bf n}^+(\eta ) = u_n^\ast(\eta )a_{\bf n}^+ (0)
  + v_n^\ast(\eta )a_{-{\bf n}} (0)
\end{equation}
where
$a_{\bf n}(0)$,
$a_{\bf n}^+(0)$,
are the initial values of the operators
$a_{\bf n}(\eta )$,
$a_{\bf n}^+(\eta )$
taken at some initial time long before the coupling became significant
and the amplification process has started, and the complex functions
$u_n$, $v_n$ satisfy the equations
\begin{equation}
 iu_n^\prime = nu_n + i(a^\prime /a) v^\ast_n \, , \quad
 iv_n^\prime = nv_n + i(a^\prime /a)u^\ast_n
\end{equation}
where
$| u_n |^2-|v_n|^2=1$
and
$u_n(0)=1$, $v_n(0)=0$.
It follows from these equations that the function
$u_n+v_n^\ast \equiv \mu_n$
obeys the equation
$\mu_n^{\prime\prime}+(n^2-a^{\prime\prime}/a)\mu_n =0$
which is precisely the equation for classical complex
$\mu$-amplitude~[2] of the gravitational-wave field.
Note that the solutions
$u_n(\eta )$, $v_n(\eta )$ to Eq.~(3) depend only on
the absolute value of the vector
${\bf n}$, $n=(n_1^2+n_2^2+n_3^2)^{1/2}$,
not its direction.  Also, these solutions are
identical for both polarizations:  they obey the same equations
with the same initial conditions.

The two complex functions $u_n$, $v_n$ restricted by one constraint
$|u_n|^2 -|v_n|^2=1$
can be parameterized by the three real functions
$r_n(\eta )$, $\phi_n(\eta )$, $\varepsilon_n(\eta )$:
\begin{equation}
u_n=e^{i\varepsilon_n}ch\, r_n \, , \qquad
v_n=e^{-i(\varepsilon_n-2\phi_n)}sh \, r_n \, .
\end{equation}
For each $n$ these functions obey the equations
\begin{equation}
r^\prime = (a^\prime /a)\cos 2\phi \, , \quad
\phi^\prime=-n-(a^\prime /a)\sin 2\phi \, cth\, 2r \, , \quad
\varepsilon^\prime = -n-(a^\prime /a) \sin 2\phi \, th\, r
\end{equation}
which can be used for an explicit calculation of
$r_n$, $\phi_n$, $\varepsilon_n$
if a time-dependent scale factor $a(\eta )$ is given.

The operators $a_{\bf n}(0)$, $a_{\bf n}^+(0)$
(${\rm Schr\ddot o dinger}$ operators) satisfy the usual commutation relations
$[a_{\bf n}(0),a_{\bf m}^+(0)]=\delta^3({\bf n}-{\bf m})$
and the same is true for the evolved operators:
$[a_{\bf n}(\eta ),a_{\bf m}^+(\eta )]=\delta^3({\bf n}-{\bf m})$.
By using Eq.~(4) the (Bogoliubov) transformation (2) can be cast in the
form
\begin{equation}
    a_{\bf n}   (\eta ) = RSa_{\bf n} (0)S^+R^+ \, , \quad
    a_{\bf n}^+ (\eta ) = RSa_{\bf n}^+ (0)S^+R^+
\end{equation}
where
\[
   S(r,\phi )=\exp \left[ r \left( e^{-2i\phi} a_{\bf n} (0)a_{-\bf n} (0)
   -e^{2i\phi}a_{\bf n}^+(0)a_{-\bf n}^+(0) \right) \right]
\]
is the unitary two-mode squeeze operator and
\[
        R(\varepsilon )=\exp
\left[ -i\varepsilon \left( a_{\bf n}^+ (0)a_{\bf n}(0)
     +  a_{-\bf n}^+ (0)a_{-\bf n}(0) \right) \right]
\]
is the unitary rotation operator.  The functions
$r_n$, $\phi_n$, $\varepsilon_n$
are called squeeze parameter, squeeze angle and rotation
angle.  (For a description of squeezed states see,
for example,~[4].)  Equations~(2), (6) demonstrate explicitely
the inevitable appearance of squeezing in the problems of this kind.
In this paper we use the presentation based on travelling waves
and two-mode squeezed states but standing waves and one-mode squeezed
states are equally good~[3].

We assume that the quantum state of the field is the vacuum state
defined by the requirement
$a_{\bf n}(0)|0> =0$ for each ${\bf n}$ and for both $s$.
In the Heisenberg picture the state of the field does not change
in time but the operators do.  The values of
$a_{\bf n}(\eta )$, $a_{\bf n}^+(\eta )$
determine all the statistical properties of the field at
the later times.  It follows from Eq.~(2) that the mean values of
$a_{\bf n}(\eta )$, $a_{\bf n}^+(\eta )$
are zero:
$<0|a_{\bf n}(\eta )|0> = 0$,
$<0|a_{\bf n}^+(\eta )|0>=0$,
but the mean values of the quadratic combinations of
$a_{\bf n}(\eta )$, $a_{\bf n}^+(\eta )$
(variances) are not zero:

\begin{eqnarray}
      <0 |a_{\bf n}(\eta )a_{\bf m} (\eta )|0>
& = & u_n(\eta) v_m(\eta )\delta ^3 ({\bf n}+{\bf m}) \nonumber \\
      <0 |a_{\bf n}^+(\eta )a_{\bf m}^+ (\eta )|0>
& = & v_n^\ast(\eta) u_m^\ast (\eta )\delta ^3 ({\bf n}+{\bf m}) \\
      <0 |a_{\bf n}(\eta )a_{\bf m}^+ (\eta )|0>
& = & u_n(\eta) u_m^\ast (\eta )\delta ^3 ({\bf n}-{\bf m}) \nonumber \\
      <0 |a_{\bf n}^+ (\eta )a_{\bf m} (\eta )|0>
& = & v_n^\ast(\eta) v_m (\eta )\delta ^3 ({\bf n}-{\bf m}) \nonumber
\end{eqnarray}
These relationships (the first two) show explicitely that the waves (modes)
with the opposite momenta are not independent. On the contrary, they
are strongly correlated which is the reason for the appearance of standing
waves.  This fact finds its reflection in the correlation functions of the
field.

To simplify the discussion of the correlation functions, we will first
ignore the tensorial indices in Eq.~(1) and consider a scalar field
\[
  h(\eta ,{\bf x)} = \int_{-\infty}^\infty d^3{\bf n}
  [a_{\bf n}(\eta )e^{i{\bf nx}} + a_{\bf n}^+(\eta )e^{-i{\bf nx}}] \, .
\]
Physically, the field
$h(\eta ,{\bf x})$
may be a scalar variable associated with the density perturbations
(see Ref.~[5] and the third paper in Ref.~[3]).
The mean value of the field $h$ is zero in every spatial point
and at every moment of time.  The variance of the field is not zero,
it can be calculated with the help of Eq.~(7):
\[
   <0|h(\eta ,{\bf x})h(\eta ,{\bf x})|0>
 = 4\pi \int_0^\infty n^2
   dn(|u_n|^2 + |v_n|^2 + u_n v_n + u_n^\ast v_n^\ast ) \, .
\]
In terms of the squeeze parameters the result can be written as
\begin{equation}
   <0|h(\eta ,{\bf x}) h(\eta ,{\bf x})|0>
=  4\pi \int_0^\infty n^2dn(ch2r_n+sh2r_n \cos 2\phi_n )
\end{equation}
(this expression includes the vacuum energy term
$4\pi\int_0^\infty n^2dn$
which should be subtracted at the end).
The variance of the field does not depend on the spatial coordinate
${\bf x}$ but does depend, in general, on time.
The function under the integral in Eq.~(8) is usually called
the power spectrum of the field:
$P(n) =n^2(ch2r_n+sh2r_n\cos2\phi_n)$.
The important property of squeezing is that, for a given time,
the function $P(n)$ is not a smooth function of
$n$ but is modulated and contains many zeros or,
strictly speeking, very deep minima.  To see this,
one can return to Eqs.~(5).  For late times, that
is, well after the completion of the amplification process,
the function $a^\prime /a$ on the right-hand side of Eqs.~(5)
can be neglected. (This is equivalent to saying that one
is considering waves that are well inside the Hubble
radius.)  At these late times, the squeeze parameter $r_n$
is not growing any more and the squeeze angle is just
$\phi_n=-n\eta -\phi_{0n}$.  Since $r_n \gg 1$ for the
frequencies of our interest~[3], the $P(n)$ can be written as
$P(n) \approx n^2e^{2r_n}\cos^2(n\eta +\phi_{0n})$.
The factor $\cos^2(n\eta+\phi_{0n})$ vanishes
for a series of values of $n$; at these frequencies the function
$P(n)$ goes to zero.  The position of zeros, as a function
of $n$, varies with time.  The similar conclusions hold for
the spatial auto-correlation function:
\[
   <0|h(\eta ,{\bf x})h(\eta ,{\bf x}+{\bf l})|0>
 = 4\pi \int_0^\infty n^2 {\sin nl \over nl}
   (ch2r_n + sh2r_n\cos2\phi_n)dn \,.
\]
The resulting expression depends on the distance between the points but not
on their coordinates. The power spectrum of this correlation function is
also modulated by the same factor $\cos^2(n\eta +\phi_{0n})$.

We return now to the tensor field (1). There is one combination of
the components $h_{ij}$ which has a special
meaning: $h(e^k)=h_{ij}e^ie^j$, where  \\
$e^k=(\sin\bar\theta\cos\bar\phi ,\sin\bar\theta\sin\bar\phi ,\cos\bar\theta)$
is an arbitrary unit vector.  The $h(e^k)$ enters the calculation of the CMBR
temperature variation seen in the
direction $e^k$ (Sachs-Wolfe effect~[6]):
\[ {\delta T \over T}(e^k) = {1\over 2} \int_0^{w_1}
   \left( {\partial h_{ij} \over \partial\eta}e^ie^j \right) dw
\]
where
$w=\eta_R-\eta$, $x^k=e^kw$, $w_1=\eta_R -\eta_E$ and $h_{ij}$ in this
formula is $a^{-1}(\eta )$ times $h_{ij}$ introduced in Eq.~(1).
For a quantized $h_{ij}$-field, the $\delta T/T$ becomes an operator:
\begin{eqnarray}
       {\delta T \over T} (e^k)
 & = & {1 \over 2} C \int_0^{w_1} dw
       \int_{-\infty}^\infty d^3{\bf n} \sum_{s=1}^2 p_{ij}^s({\bf n})
        e^ie^j \{ [\alpha_n^s a_{\bf n}^s (0)
      + \beta_n^s a_{-\bf n}^{s+}(0)] e^{in_ke^kw} \nonumber \\
 & + & [\alpha_n^{s\ast} a_{\bf n}^{s+} (0)
     + \beta_n^{s\ast}a_{-\bf n}^s(0)] e^{-in_k x^k w} \} \nonumber
\end{eqnarray}
where
$\alpha_n^s(\eta) \equiv (u_n^s/a)^\prime$,
$\beta_n^s(\eta) \equiv (v_n^s/a)^\prime$.
The mean value of $\delta T/T$ is zero
while the variance of the expected temperature
fluctuations can be written as
\begin{eqnarray}
    <0| {\delta T \over T} (e^k){\delta T \over T} (e^k)|0>
& = & {1 \over 4} C^2 \int_0^{w_1} dw \int_0^{w_1} d \bar w
    \int_{-\infty}^\infty d^3{\bf n} \cos(n_ke^k\xi) \nonumber \\
& \times & \pi^1({\bf n},e^k) f(n,w,\bar w )
\end{eqnarray}
where $\xi =w-\bar w$ and
\begin{eqnarray}
   \pi^1 ({\bf n},e^k) & \equiv & (p_{ij}^1 ({\bf n})e^ie^j)^2
                         + (p_{ij}^2 ({\bf n})e^ie^j)^2 , \nonumber \\
   f(n,w,\bar w ) & \equiv & \alpha_n(w)\alpha_n^\ast (\bar w )
                    + \beta_n^\ast (w)\beta_n(\bar w )
                    + \alpha_n(w)\beta_n (\bar w )
                    + \beta_n^\ast (w)\alpha_n^\ast (\bar w ), \nonumber \\
   \alpha_n^1 & = & \alpha_n^2 \equiv \alpha_n \, , \quad
                    \beta_n^1 = \beta_n^2 \equiv \beta_n \, . \nonumber
\end{eqnarray}

The integration over the variables $\varphi$, $\theta$ in Eq.~(9) allows
one to reduce this formula to
\begin{eqnarray}
     <0| {\delta T \over T}(e^k) {\delta T \over T} (e^k) |0>
   = C^2 8\pi \int_0^{w_1} dw \int_0^{w_1} d\bar w \int_0^{\infty} n^2
     W_1(n\xi )f(n,w,\bar w) dn
\end{eqnarray}
where
\begin{eqnarray}
     W_1(n\xi )
   = (\pi /2)^{1/2} (n\xi)^{-5/2} {\it J}_{5/2} (n\xi)\, . \nonumber
\end{eqnarray}
The term
$W_1(n\xi)$ depends on the interval between the points but
not on the direction of sight.  Thus, variancies seen in all
directions $e^k$ are the same.  They are also position
independent as for $x^k = e^kw+x_0^k$ the coordinates $x_0^k$ of
the observer drop out of the final result.

We will now turn to the derivation of the angular correlation function \\
$<0|\delta T/T(e_1^k)\delta T/T(e_2^k)|0>$ where $e_1^k$ and
$e_2^k$ are two different unit vectors.  The general formula for
this function can be written as
\begin{eqnarray}
    <0|{\delta T \over T} (e_1^k){\delta T \over T} (e_2^k)|0>
& = & {1\over 4} C^2 \int_0^{w_1} dw \int_0^{w_1} d\bar w
    \int_{-\infty}^\infty d^3{\bf n} \cos(n_i\zeta^i) \nonumber \\
& \times & \pi^2 ({\bf n},e_1^k,e_2^k)f(n,w, \bar w)
\end{eqnarray}
where $\zeta^i =e_1^iw-e_2^i \bar w$ and
\begin{eqnarray}
         \pi^2({\bf n},e_1^k,e_2^k)
  \equiv (p_{ij}^1({\bf n})e_1^ie_1^j)(p_{lm}^1({\bf n})e_2^le_2^m)
       + (p_{ij}^2(n)e_1^ie_1^j)(p_{lm}^2({\bf n})e_2^le_2^m). \nonumber
\end{eqnarray}
The integration over the variables $\varphi ,\theta$ in Eq.~(11)
reduces this formula to
\begin{eqnarray}
    <0|{\delta T \over T} (e_1^k){\delta T \over T} (e_2^k)|0>
 =  C^2 8\pi\int_0^{w_1} dw \int_0^{w_1} d\bar w \int_0^\infty n^2
    W_2(n\zeta , \cos\delta ) f(n,w, \bar w ) dn \nonumber
\end{eqnarray}
\begin{eqnarray}
\end{eqnarray}
where $\zeta =(w^2-2w\bar w \cos\delta +\bar w^2)^{1/2}$,
$\delta$ is the angle between the two directions of observation,
$\cos\delta = e_1^1e_2^1+e_1^2e_2^2+e_1^3e_2^3$, and
\begin{eqnarray}
      W_2(n\zeta ,\cos\delta )
& = & {1\over2}(3\cos^2\delta -1)(\pi /2)^{1/2} (n\zeta )^{-5/2}
    J_{5/2}(n\zeta ) \nonumber \\
& + & \cos\delta(\cos^2\delta -1)(nw)(n\bar w)(\pi /2)^{1/2}
      (n\zeta )^{-7/2} J_{7/2}(n\zeta ) \nonumber \\
& + & {1\over 8}(cos^2 \delta -1)^2(nw)^2(n\bar w )^2
      (\pi /2)^{1/2} (n\zeta )^{-9/2} J_{9/2}(n\zeta ).
\end{eqnarray}
Expression (12) depends only on $\cos\delta$ and, hence, the
correlation function is rotationally symmetric.
In the limit $\cos\delta =1$, the parameter $\zeta$ goes over
into $\xi$ and Eq.~(12) coincides with Eq.~(10).

Expression (12) gives the angular correlation function in the
general and universal form.  It can be used with arbitrary
functions $\alpha_n(w),\beta_n(w)$, that is, it is applicable for
arbitrary (not necessarily inflationary) cosmological models generating
squeezed gravitational
waves.  The remaining integrations in Eq.~(12) assign concrete
numerical values to the correlations attributed to different
separation angles $\delta$, but they do not change the general
angular pattern represented by the function
$W_2(n\zeta ,\cos\delta )$.  Consistency with the data of the
COBE-type observations may lead to the determination of the
functions $\alpha_n(w),\beta_n(w)$ and, eventually, to the
knowledge of the expansion rate of the early universe.  This will
be a subject of a separate discussion.  The implications of the
COBE observations for inflationary models are under active
analysis (see, for instance, a recent paper [7] and references
therein).  Some new results based on the correlation function
(12), (13) have been derived in [8].

\bigskip

This work was supported in part by NASA grant NAGW~2902 and NSF
grant 89-22140.

\end{document}